# PhyloProfile v2: Scalable Exploration of Multi-layered Phylogenetic Profiles via Dimensionality Reduction

Vinh Tran[1,*] and Ingo Ebersberger[1,2,3*]

[1]Department for Applied Bioinformatics, Institute of Cell Biology and Neuroscience, Goethe University, 60438 Frankfurt am Main, Germany [2]Senckenberg Biodiversity and Climate Research Centre (BiK-F), 60325 Frankfurt am Main, Germany [3]LOEWE Centre for Translational Biodiversity Genomics (TBG), 60325 Frankfurt am Main, Germany

*To whom correspondence should be addressed.

## Abstract

**Motivation:** Phylogenetic profiles—presence-absence patterns of genes across taxa—are rich information sources for inferring the evolutionary history of genes and gene families. When aggregated across many genes, these profiles can reveal coevolutionary patterns, supporting the prediction of gene functions and interactions. With rapidly growing numbers of sequenced genomes, phylogenetic profiles now routinely encompass thousands of genes and taxa. Existing software fall short in enabling interactive visualization, exploration, and analysis of such large datasets. We present PhyloProfile v2, a comprehensive overhaul of the original PhyloProfile software. This new version introduces major performance improvements along with novel features designed for more efficient data exploration. Notably, PhyloProfile v2 integrates dimensionality reduction techniques to visualize phylogenetic profiles in interactive 2D or 3D space, offering an intuitive overview even for massive datasets. Furthermore, the platform enables seamless transitions from large-scale analyses—spanning millions of orthology relationships—to detailed comparisons of protein feature architectures between specific orthologs. PhyloProfile v2 thus provides a versatile and scalable solution for evolutionary and functional genomics research.

**Availability and implementation:** PhyloProfile v2 is available as an R package at Bioconductor https://doi.org/doi:10.18129/B9.bioc.PhyloProfile. The open-source code and documentation are provided under MIT license at https://github.com/BIONF/PhyloProfile

**Contact:** tran@bio.uni-frankfurt.de or ebersberger@bio.uni-frankfurt.de

## 1 Introduction

Phylogenetic profiles capture the presence/absence patterns of homologs across species. They provide insights into the evolution of gene sets such as lineage-specific gene gains and losses, horizontal gene transfer, and signals of co-evolution (Pellegrini *et al.* 1999; Birikmen *et al.* 2021; Linard *et al.* 2021; Muelbaier *et al.* 2024; Rossier *et al.* 2024; Tran *et al.* 2024). Meanwhile, phylogenetic profiles are often supplemented with additional information layers. Examples are pairwise sequence similarities (Enault *et al.* 2004; Tabach *et al.* 2013; Bloch *et al.* 2020), information about the evolutionary relationships of the subsumed genes (Deng, Hernández-Plaza and Huerta-Cepas 2023; Rossier *et al.* 2024), or measures describing the extent to which proteins differ in their feature architectures, where typical features represent Pfam and SMART domains, transmembrane domains, signal peptides, and low complexity regions (Birikmen *et al.* 2021; Dosch *et al.* 2023; Iruegas *et al.* 2023; Altenhoff *et al.* 2024). Several tools can visualize multi-layered phylogenetic profiles. Most of them are designed to only deal with input data from specific online resources, e.g., PhyloPro2.0 (Cromar *et al.* 2016), OrthoInspector (Nevers *et al.* 2019), TreeProfiler (Deng, Hernández-Plaza and Huerta-Cepas 2023) or Matreex (Rossier *et al.* 2024). PhyloProfile (Tran, Greshake Tzovaras and Ebersberger 2018) allows users to work with their own orthology assignments in custom taxon sets. Independent of the data source, all tools share the same limitations. Both their computational capacities and current strategies for visualizing and exploring phylogenetic profiles fall short when applied to datasets of contemporary scale comprising thousands of genes and species (Langschied *et al.* 2024; Tran *et al.* 2025).

Here we present PhyloProfile v2, a tool for the interactive visualization and exploration of phylogenetic profiles that scales to millions of pairwise orthology relationships. In addition to significant performance improvements, PhyloProfile v2 has entirely reworked routines to cluster either genes or taxa according to the information in the phylogenetic profiles. As a key innovation, we introduce dimensionality reduction techniques (PCA, t-SNE, UMAP) as novel methods for the interpretation of large collections of phylogenetic profiles.

## 2 Novel features and functionalities

### 2.1 Multi-scale phylogenetic profile plot

PhyloProfile v2 includes interactive controls for adjusting the granularity of phylogenetic profile visualizations in real time. To support broader context, a static overview plot distills the entire dataset into a clear and compact summary (Figure 1A). This visualization is suitable for detecting major evolutionary signals, such as the identification of genes of the same evolutionary age (Domazet-Lošo, Brajković and Tautz 2007), coordinated gene losses on individual taxonomic lineages (Langschied *et al.* 2023), or the identification of whole genome duplication events indicated by lineage-specific presence of co-orthologs. A rectangular selection tool can be used to zoom in to individual regions of this overview plot. The selection is then displayed as an interactive taxon-gene matrix for higher resolving analyses (Figure 1B). Clicking a dot in this matrix displays accessory information, including taxonomic details with links to the NCBI taxonomy database, and, if provided, protein sequences as well as a feature architecture comparison between the two orthologs (Figure 1C). The dynamic adjustment of the resolution level enables the analysis to easily scale from the full dataset comprising several million data points down to a subset of selected orthologs, and even to a pairwise comparison of two proteins.

## 2.2 Dimensionality reduction plots

With increasing numbers of genes and/or species, taxon-gene matrices quickly grow to an extent that they can no longer be displayed with a sufficiently high resolution within a browser window. Moreover, the linear ordering of taxa based on their taxonomic relationships renders it hard to identify distantly related species that (convergently) share similar phylogenetic profiles, or species that deviate from the expected representation of genes given the taxonomic clade they reside in. In PhyloProfile v2, we introduce dimensionality reduction techniques to place taxa either in a 2D or 3D space (Figure 1D). The spatial position of each taxon is determined by the vector capturing the presence-absence pattern of the investigated genes. Species can be grouped into higher-order taxonomic groups, e.g., class or phylum, and the group assignment of each species is indicated by the dot color. For generating these plots, we leverage three widely used dimensionality reduction techniques: UMAP (McInnes, Healy and Melville 2020), t-SNE (Maaten and Hinton 2008) and PCA (Gewers *et al.* 2021). After clustering, users can select the taxa directly in the plot (Figure 1E), triggering the display of corresponding phylogenetic profiles in a conventional matrix format (see Figure 1B) for further analysis. Alternatively, the underlying information can be displayed in a tabular format (Figure 1F). This novel way of representing phylogenetic profiles is also applicable to the genes, which provides an alternative method for phylogenetic profiling.

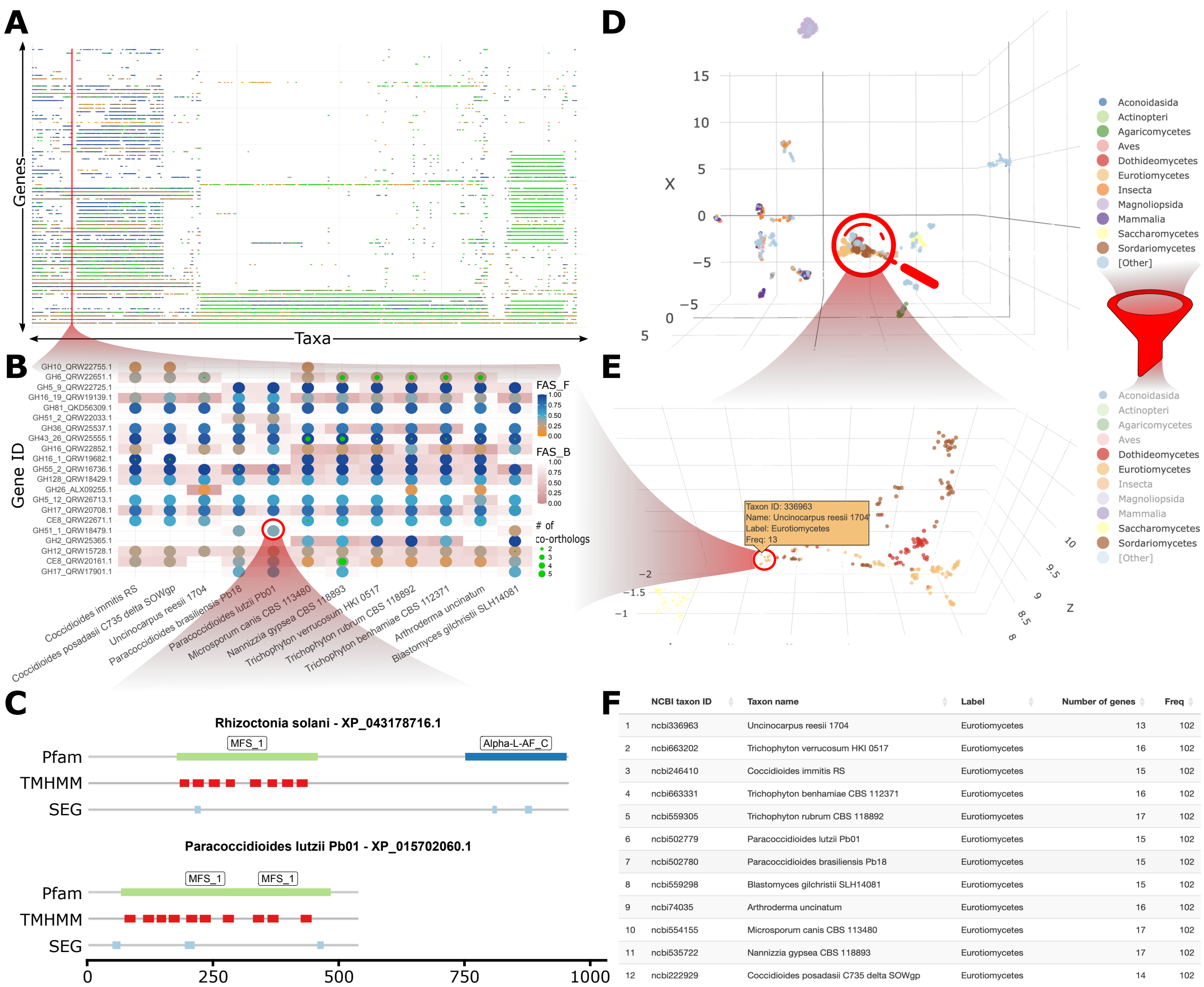

| | NCBI taxon ID | Taxon name | Label | Number of genes | Freq |
|---|---|---|---|---|---|
| 1 | ncbi336963 | Uncinocarpus reesii 1704 | Eurotiomycetes | 13 | 102 |
| 2 | ncbi663202 | Trichophyton verrucosum HKI 0517 | Eurotiomycetes | 16 | 102 |
| 3 | ncbi246410 | Coccidioides immitis RS | Eurotiomycetes | 15 | 102 |
| 4 | ncbi663331 | Trichophyton benhamiae CBS 112371 | Eurotiomycetes | 16 | 102 |
| 5 | ncbi559305 | Trichophyton rubrum CBS 118892 | Eurotiomycetes | 17 | 102 |
| 6 | ncbi502779 | Paracoccidioides lutzii Pb01 | Eurotiomycetes | 15 | 102 |
| 7 | ncbi502780 | Paracoccidioides brasiliensis Pb18 | Eurotiomycetes | 15 | 102 |
| 8 | ncbi559298 | Blastomyces gilchristii SLH14081 | Eurotiomycetes | 15 | 102 |
| 9 | ncbi74035 | Arthroderma uncinatum | Eurotiomycetes | 16 | 102 |
| 10 | ncbi554155 | Microsporum canis CBS 113480 | Eurotiomycetes | 17 | 102 |
| 11 | ncbi535722 | Nannizzia gypsea CBS 118893 | Eurotiomycetes | 17 | 102 |
| 12 | ncbi222929 | Coccidioides posadasii C735 delta SOWgp | Eurotiomycetes | 14 | 102 |

*Figure 1. Data visualization across scales in PhyloProfile v2. The figure shows the profiles of 76 plant cell-wall degrading enzymes across 1,331 eukaryotes (Tran et al. 2024). (A) Conventional taxon-gene matrix providing an overview of the full data. Each dot indicates the presence of a gene in a taxon. In the example, dot color informs about the feature architecture similarity ranging from blue (high similarity) to orange (low similarity). Green inlays denote the presence of co-orthologs. A rectangular selection tool allows to represent individual sections in a fully interactive taxon-gene matrix. A mouse click on a dot in this matrix provides further information about the respective gene pair. Moreover, the respective protein feature architectures can be displayed (C) if this information was provided as input. (D) Representation of phylogenetic profiles in 3D using UMAP. Each dot represents a species, the spatial placement of a dot is determined by the presence-absence vector across the 76 enzymes, and the dot color represents the taxonomic label listed in the legend (see main text for details). (E) A zoom in into the plot shown in (D) with a focus on individual eurotiomycetes* featuring a markedly low number of plant cell-wall degrading enzymes (red circle). Hovering over individual dots provides accessory information such as the species name, NCBI taxonomy id, taxonomic label, and numbers of target genes represented by an ortholog in this species. A click on a dot adds this *information to a table (F) and the* species to a conventional taxon gene matrix for further analysis (see B).

### 2.3 Protein feature architecture comparison

Comparing feature architectures of related proteins reveals signals for a functional diversification or issues with the gene annotation (Dosch *et al.* 2023). We have implemented a redesigned protein architecture plot in PhyloProfile v2 (see Figure 1C). Users can now adjust the plot, for example, by excluding a specific feature class, toggling the display of E-value or bit-score of predicted features, filtering domains or linearizing overlapping domains using their E-values, bit-scores, or by selecting the path that maximizes the feature architecture similarity score (see (Dosch *et al.* 2023)). Furthermore, links to InterPro (Blum *et al.* 2025) for Pfam domains and SMART Database for SMART domains are provided, facilitating more informed predictions regarding the protein's functional annotation.

### 2.4 Performance improvement

After optimizing data processing and graphical rendering, we benchmarked PhyloProfile v2 using simulated datasets. We generated phylogenetic profiles with increasing numbers of taxa or genes by keeping one of the two parameters constant at 1,000 items while varying the other parameter across 1,000, 2,000, 3,000, 4,000, 5,000, 10,000, and 15,000 items. The resulting datasets comprise between 500,000 and 7.5 million ortholog pairs. PhyloProfile v2 runs 2- to 5-fold faster than PhyloProfile v1.18, the last official release, and the run time improvement becomes more pronounced as dataset sizes increase. The benchmarks were performed on a laptop with an Apple M1 processor and 16 GB of RAM. The complete performance benchmark is available on the Github Wiki page https://github.com/BIONF/PhyloProfile/wiki/Performance-test.

### 2.5 Flexibility in data handling and visualization

PhyloProfile v2 is a stand-alone software. Formatting requirements for the input data are provided in the software manual. By default, species are ordered using the NCBI taxonomy (Schoch *et al.* 2020), but users can provide alternative taxonomies. Numerous options allow users to adjust the presentation of the phylogenetic profiles. For example, (i) dynamically filtering of phylogenetic profiles using custom cutoff values for, e.g., the FAS score or the maximum number of co-orthologs, (ii) dynamically adjusting the taxonomic resolution in the profile plot from the strain level to summarizing species on the superkingdom level, (iii) displaying selected genes and taxa in a custom plot, and (iv) modifying the taxon labeling in the dimensionality reduction plots by setting the taxonomic rank of the labels or manually grouping labels into higher taxonomic ranks. Download options include the plots as images, tables in csv format, amino acid sequences in FASTA format, as well as filtered data in the phyloprofile input format. This makes it straightforward to integrate the management, exploration, and analysis of large-scale phylogenetic profiles into workflows investigating the evolutionary histories of taxa, their genes, and the functions they convey.

## Acknowledgements

The authors thank Felix Langschied for valuable discussion and for proofreading this manuscript.

## Funding

The research was funded by the Landes-Offensive zur Entwicklung Wissenschaftlich-ökonomischer Exzellenz (LOEWE) Program of the Hessian Ministry of Higher Education, Research, Science and the Arts through the LOEWE Centre for Translational Biodiversity Genomics (LOEWE- TBG).

*Conflict of Interest:* none declared.